\documentstyle[preprint,aps]{revtex}
\begin{document}
\draft

\title{Effect of Iodine Doping on Bi$_{2}$Sr$_{2}$Ca$_{1}$Cu$_{2}$O$_{x}$:\\
Charge Transfer or Interlayer Coupling?}

\author{Jian Ma,$^{a}$ P.Alm\'{e}ras,$^{b}$ R.J.Kelley,$^{a}$ H.Berger,$^{b}$\\
G. Margaritondo,$^{b}$
A.Umezawa,$^{c}$ Marvin L. Cohen$^{d}$ and M.Onellion$^{a}$}
\address{$^{a}$Department of Physics, University of Wisconsin-Madison,
Madison, WI 53706\\
$^{b}$Institut de Physique Appliqu\'{e}e, Ecole Polytechnique F\'{e}d\'{e}rale,
CH-1015 Lausanne, Switzerland\\
$^{c}$Applied Superconductivity Center, University of Wisconsin-Madison,
Madison, WI 53706\\
$^{d}$Department of Physics, University of California at Berkeley,\\
and Materials Sciences Division, Lawrence Berkeley Laboratory,
Berkeley, CA 94720}

\date{\ \ Received    }

\maketitle

\begin{abstract}
A comparative study has been made of iodine-intercalated
Bi$_{2}$Sr$_{2}$Ca$_{1}$Cu$_{2}$O$_{x}$ single crystal and 1 atm O$_{2}$
annealed Bi$_{2}$Sr$_{2}$Ca$_{1}$Cu$_{1}$O$_{x}$ single crystal using
AC susceptibility measurement, X-ray photoemission (XPS) and angle-resolved
ultraviolet photoemission spectroscopy (ARUPS). AC susceptibility measurement
indicates that O$_{2}$-doped samples studied have T$_{c}$ of 84 $^{o}$K,
whereas T$_{c}$ of Iodine-doped samples studied are 80 $^{o}$K. XPS Cu 2p
core level data establish that the hole concentration
in the CuO$_{2}$ planes are essentially the same for these two kinds of
samples.  ARUPS measurements show that electronic structure of the normal
states near the Fermi level has been strongly affected by iodine intercalation.
We conclude that the dominant effect of iodine doping is to alter the
interlayer coupling.
\end{abstract}


%\section{Introduction}
%\label{sec:level1}

The normal state properties of high T$_{c}$ cuprate superconductors continues
to
be of much interest since probing the nature of the normal state may
provide the key to understanding the mechanism of high-T$_{c}$
superconductivity.\cite{Anderson1} Since the report that iodine has been
successfully intercalated between BiO bilayers of the
Bi$_{2}$Sr$_{2}$Ca$_{1}$Cu$_{2}$O$_{x}$, \cite{Xiang1}
many studies have been made of its crystal structure,
\cite{Xiang1,Xiang2,Qiu,Chenevier,Kijima} transport properties
\cite{Xiang2,Nakashima,Xiang3,Pooke}
and electronic properties.\cite {Fukuda,Pooke,Choy} A key unanswered question
is how iodine intercalation alters the superconducting properties  of
intercalated Bi$_{2}$Sr$_{2}$Ca$_{1}$Cu$_{2}$O$_{x}$.
By using transmission electron microscopy(TEM) and X-ray diffraction,
Xiang et al\cite{Xiang2,Xiang3} concluded that iodine is epitaxially
intercalated between the Bi-O bilayers with CuO$_{2}$ planes unaffected.
They found that metallic CuO$_{2}$ planes in the normal
state above T$_{c}$ are unaffected by the intercalation; however, the c-axis
electrical resistivity is changed from semiconducting to metallic-like
behavior. They attributed the reduction of T$_{c}$ for
IBi$_{2}$Sr$_{2}$Ca$_{1}$Cu$_{2}$O$_{x}$ to the change
of interlayer coupling, but did not rule out a smaller effect on  T$_{c}$
due to charge transfer from the iodine layers to the CuO$_{2}$ layers.
By contrast,
based on X-ray photoemission(XPS) and Hall coefficient measurements,
Pooke et al\cite{Pooke} and Choy et al\cite{Choy} concluded that the dominant
cause of changes in T$_{c}$ is charge transfer from the iodine to the
CuO$_{2}$ planes. Whether T$_{c}$ is controlled by in-plane carrier
density or by
interlayer coupling has been an important issue for a long time in the
high T$_{c}$ superconductivity community.\cite{Anderson2} Our approach is to
perform a
comprehensive and comparative study of iodine-intercalated Bi-2212 and
1 atm O$_{2}$ annealed Bi-2212. We concluded that it is the change of
interlayer
coupling due to the iodine intercalation that dominates
the change of T$_{c}$ between the two kinds of samples that have been studied.
Using synchrotron radiation angle-resolved photoemission, we found that
the electronic structure along one of the main symmetry directions,
specifically $\Gamma$-X(corresponding to the Bi-O bond axis in real space),
has been strongly affected by this change of
interlayer coupling associated with iodine intercalation.

%\section{Experimental}

The stage one IBi$_{2}$Sr$_{2}$Ca$_{1}$Cu$_{2}$O$_{x}$ samples were
prepareed using the method reported
by Xiang et al.\cite{Xiang1,Xiang2,Xiang3} The samples were characterized
by X-ray diffraction, micro-probe (Cameca CAMEBAX SX-50) and
resistivity measurements.

The AC susceptibility measurements were obtained using a Lakeshore 7000
series susceptometer/magnetometer. The temperature step size has
been chosen so that the determination of T$_{c}$ was accurate to 0.2 K.

The X-ray photoemission spectroscopy(XPS) measurements were performed in
Lausanne Centre de Spectromicroscopie, Switzerland, using the SCIENTA
ESCA-300 system which is equipped with a rotating anode Al K$\alpha$ radiation
source, a seven-quartz-crystal monochromator and a concentric hemispherical
analyser. The overall energy resolution of the system for this measurement
was 340 meV. The chamber pressure during the measurement was
2 x 10$^{-10}$ torr.

The angle-resolved ultraviolet photoemission measurements were conducted
using the
four meter normal incidence monochromator at the Synchrotron Radiation Center
in Wisconsin, and a Vacuum Science Workshop hemispherical angle-resolved
analyser which has an acceptance full angle of two degrees. The experiment
was performed at room temperature. The energy resolution of 140 meV was
determined by measuring the Fermi edge of a gold film. The norminal chamber
pressure during the measurement was 1.7 x 10$^{-10}$ torr.

In both XPS and ARUPS measurements, the samples were placed in a load lock
chamber with a base pressure lower than 5 x 10$^{-9}$ torr, transferred into
the
main chamber, and cleaved under an ultra-high vacuum of 2 x 10$^{-10}$. In
this way we made certain that both iodine-doped samples and O$_{2}$
annealed samples retained their original stoichiometries.

We confirmed the sample orientation using low energy electron diffraction.
Our measurements along $\Gamma$-X in reciprocal space correspond to
the Bi-O bond axis direction in real space, while the $\Gamma$-M direction
in reciprocal space corresponds to the Cu-O-Cu bond axis direction in
real space.

In order to separate the effects of charge transfer and interlayer
coupling, we carefully altered the oxygen stoichiometry so that
the Cu$^{+1}$:Cu$^{+2}$ ratio obtained from XPS data was the same for the two
types of samples. By doing so, we intended to determine the effects of
interlayer coupling, without differences in charge transfer complicating
the interpretation.

%\section{Results and Discussion}

Fig.\ \ref{Copper} illustrates Cu 2p$_{3/2}$ XPS core-level spectra for
the B-2212 single crystals with 1 atm O$_{2}$ annealing at
650 $^{o}$C for 16 hours(a) and with iodine intercalation(b).We used one
atmosphere O$_{2}$ annealing to obtain as closely as possible the same
carrier density for samples with and without iodine intercalation. From the
spectrum one can observe a main peak at about 933 eV corresponding to the
well-screened core hole final state of the 2p$^{5}$3d$^{10}$L, where
L indicates a hole
on the ligand oxygen, and a week broad satellite centered at about 945 eV
corresponding to a poorly-screened final state of the 2p$^{5}$3d$^{9}$
configuration.
We deconvoluted the main peak into three components. The two lower binding
energy components comprise the Cu$^{+1}$. The satellite and the rest of
higher binding energy components comprise Cu$^{+2}$.\cite{Tranquada,Fukuda}
Using this, the ratio of Cu$^{+1}$:Cu$^{+2}$ for two kinds of samples are
found to be $0.87\pm0.03$ for I-doped sample
and $0.92\pm0.03$ for O$_{2}$ annealed sample. These ratios are
the same within experimental accuracy. The data indicates the
carrier density in CuO$_{2}$ planes are the same for the I-doped sample and
O$_{2}$ annealed sample.

Fig.\ \ref{AC} illustrates the AC susceptibility of a Bi 2:2:1:2 single crystal
after 16 hours 1 atm O$_{2}$ annealing at 650 $^{o}$C and that of Bi 2:2:1:2
single
crystal after iodine intercalation. The onset T$_{c}$ is 84 $^{o}$K for
O$_{2}$ annealed
sample and 80 $^{O}$K for the iodine-intercalated sample. The T$_{c}$ of our
as-grown samples is 90 $^{o}$K. It is clear that there is a difference of
4 $^{o}$K
in their transition temperatures for 1 atm O$_{2}$ annealed sample and I-doped
sample.

Bi 5d$_{3/2}$ and 5d$_{5/2}$ spectra for Bi-2212 with O$_{2}$ annealing and
with iodine
intercalation are shown in Fig.\ \ref{Bimuth}. We determined that the
Bi 5d$_{3/2}$ and Bi 5d$_{5/2}$ levels have the same binding energies,
28.7 eV and 25.7 eV, respectively for the two kinds of samples . These
results are consistent with  Cu 2p$_{3/2}$ data, because we would expect
no shift of corresponding core level binding energies since there is no
difference in hole doping.

The above XPS Cu(2p) core level spectra, AC susceptibility data, Bi(5d)
core level spectra  establish that the difference in T$_{c}$
between the two types of samples studied cannot be explained in terms
of differences in hole concentration.

Xiang et al\cite{Xiang3} interpreted the observed change in T$_{c}$ using
two models: the Wheatley-Hsu-Anderson(WHA) model\cite{Wheatley} and, as
they modified in Ref.9, the Ihm-Yu model.\cite{Ihm}
They found that the depression of T$_{c}$ can be understood if one assumes
that the next-nearest CuO$_{2}$ plane coupling is essentially eliminated
by iodine intercalation; therefore, the interlayer hopping matrix element
for the next-nearest-neighbor CuO$_{2}$ planes is greatly reduced.
Such an interpretation is consistent with high pressure experiments that
report $dT_{c}/dP > 0 $.\cite{Berkley}
We studied whether there was any indication of change of interlayer
coupling from the point of view of electronic structure.
Specifically, we measured
the band structure. The spectra illustrated in Fig.\ \ref{gammax} were
taken  along the $\Gamma$-X direction using a photon energy of  21 eV .
For the iodine-intercalated sample Fig.\ \ref{gammax}, we observe a band
dispersing
towards the Fermi level from 160 meV below E$_{f}$ at $\theta$ of 10 degrees.
The band crosses the Fermi level at $\theta$ = 16 degree and remains
at the Fermi level for another two degrees. After crossing the Fermi level,
the trailing edge of this band is still visible for $\theta$ as large as
22 degrees. By contrast, for an oxygen annealed sample, the band crosses
the Fermi level at $\theta$ = 12 degree.\cite{Olson,Kelley}
The experimental dispersion relations for both iodine
doped and O$_{2}$ annealed samples are shown in Fig.\ \ref{dispersion}. The
amount of dispersion is quite different for two
types of samples. The iodine intercalated samples exhibit a reduced dispersion,
which is about 1.7 times less than that of the oxygen annealed sample,
and a larger Fermi wave vector. Band structure calculations suggest
that this band is mostly of CuO character.\cite{Freedman}
The data indicate that the change
of interlayer coupling has altered band structure in the $\Gamma$-X direction.

We performed similar measurements in the $\Gamma$-M direction. We found no
significant difference in the dispersion or Fermi wave vector between
samples.

%\section{Conclusion}

In summary, our comparative study of iodine-intercalated Bi-2212
single crystal and O$_{2}$ annealed Bi-2212 has found that the
T$_{c}$ is significantly
affected by interlayer coupling effects rather than solely by hole carrier
concentrations. Our data have established that the electronic structure of
the normal state is dependent on the strength of the
interlayer coupling.

\vspace{.8in}

%\noindent ACKNOWLEDGEMENT

One of us (J.M.) would like to thank Tian Zhang for her assistance in
programming for data analysis. Financial support by the U.S. National
Science Foundation(directly and through the Synchrotron Radiation Center
in Wisconsin), by the Wisconsin Alumni Research Foundation, by
the Fonds National Suisse de la Recherche Scientifique and by the Ecole
Polytechnique F\'{e}d\'{e}rale de Lausanne, is gratefully acknowledged.
M.L.C acknowledges helpful conversations with Alex Zettl and support
from NSF Grant No. DMR91-20269 and the Director, Office of Energy Research,
Office of Basic Energy Siciences, Materials Sciences Division of the U.S.
Department of Energy under Contract No.DE-AC03-76SF00098.
\eject

\begin{figure}
\caption{The Cu(2p$_{3/2}$) XPS core level spectra for
iodine-intercalated Bi-2212 single crystal(a), and for 1 atm O$_{2}$
annealed Bi-2212(b).}
\label{Copper}
\end{figure}

\begin{figure}
\caption{The AC susceptibility data for iodine-intercalated Bi-2212 single
crystal(a) and the data for 1 atm O$_{2}$ annealed Bi-2212(b). The onset
T$_{c}$'s are 80 $^{o}$K and 84 $^{o}$K, respectively.}
\label{AC}
\end{figure}

\begin{figure}
\caption{The Bi(5d) UPS core level spectra for iodine-intercalated
Bi-2212 single crystal(a), and for 1 atm O$_{2}$ annealed Bi-2212(b).}
\label{Bimuth}
\end{figure}

\begin{figure}
\caption{The angle-resolved photoemission spectra of iodine-intercalated
Bi-2212 taken at room temperature along the $\Gamma$-X directions.
The photon envery h$\nu$ is 21 eV.}
\label{gammax}
\end{figure}

\begin{figure}
\caption{The dispersion relations of the same band along $\Gamma$-X for
iodine-intercalated Bi-2212(a), and for 1 atm O$_{2}$ annealed Bi-2212(b). Both
curves are experimental ones.}
\label{dispersion}
\end{figure}

\end{document}